\documentclass[%
 final,reprint,altaffilletter,%
 showpacs,%
 amssymb, amsmath,%
 aps,cha, 
 noeprint,
 nolongbibliography,superscriptaddress
]{revtex4-1}

\usepackage{amsfonts}
\usepackage{graphicx}%
\usepackage[colorlinks=true,linkcolor=blue,citecolor=blue,urlcolor=black,breaklinks=true]{hyperref}%

\def\apj{Astrophys.~J.}
\def\apjl{Astrophys.~J.~Lett.}

\def\apss{Astrophys.~Space~Sci.}

\def\aap{Astron.~Astrophys.}

\def\pre{Phys.~Rev.~E}
\def\prl{Phys.~Rev.~Lett.}

\def\pasj{Publ.~Astron.~Soc.~Japan}

\def\ssr{Space~Sci.~Rev.}

\def\nat{Nature}

\def\grl{Geophys.~Res.~Lett.}

\def\jgr{J.~Geophys.~Res.}
\def\jgra{J.~Geophys.~Res. (Space Physics)}

\def\physscr{Phys.~Scr.}

\def\phfl{Phys.~Fluids}  
          
\def\phpl{Phys.~Plasmas}    
\def\ppcf{Plasma~Phys.~Contr.~Fusion}  
\def\jplph{J.~Plasma~Phys.}  
  
\def\jpsj{J.~Phys.~Soc.~Japan}

\def\cejph{Central~Eur.~J.~Phys.}  
\def\jcap{J.~Cos.~Astropart.~Phys.}  
  
\def\plphr{Plasma~Phys.~Rep.}

\begin{document}
\title[Electrostatic waves in an e-p pair plasma with suprathermal electrons]{Electrostatic solitary waves in an electron-positron pair plasma\\ with suprathermal electrons}
\author{A.~Danehkar}
\thanks{E-mail address:
\href{mailto:ashkbiz.danehkar@cfa.harvard.edu}{ashkbiz.danehkar@cfa.harvard.edu}; \\Present address: Harvard-Smithsonian Center for Astrophysics, 60 Garden Street, Cambridge, MA 02138, USA}
\affiliation{Centre for Plasma Physics, Queen's University Belfast, Belfast BT7 1NN, UK}
\affiliation{Department of Physics and Astronomy, Macquarie University, Sydney, NSW 2109, Australia}

\begin{abstract}
The nonlinear propagation of electrostatic solitary waves is studied in a collisionless electron-positron pair plasma consisting of adiabatic cool electrons, mobile cool positrons (or electron holes), hot suprathermal electrons described by a $\kappa$ distribution, and stationary ions. The  linear dispersion relation derived for electrostatic waves demonstrates a weak dependence of the phase speed on physical conditions of positrons in appropriate ranges of parameters. The Sagdeev's pseudopotential approach is used to obtain the existence of electrostatic solitary wave structures, focusing on how their characteristics depend on the physical conditions of positrons and suprathermal electrons. Both negative and positive polarity electrostatic solitary waves are found to exist in different ranges of Mach numbers. As the positrons constitute a small fraction of the total number density, they slightly affect the existence domains. However, the positrons can significantly change the wave potential at a fixed soliton speed. The results indicate that the positive potential can greatly be grown by increasing the electron suprathermality (lower $\kappa$) at a fixed true Mach number. It is found that a fraction of positrons maintain the generation of positive polarity electrostatic solitary waves in the presence of suprathermal electrons in pair plasmas.
\end{abstract}
\volumeyear{year}
\volumenumber{number}
\issuenumber{number}
\eid{identifier}
\received[Received: ]{19 August 2017}

\accepted[Accepted: ]{11 October 2017}

\published[Published: ]{26 October 2017}

\maketitle

\section{Introduction\label{sec:introduction}}

Electron-positron (e-p) pair plasmas are present in many astrophysical
environments such as the solar wind \cite{Clem1996,Protheroe1982,Moskalenko1998,Clem2000,Adriani2009,Adriani2011}, the Earth's
magnetosphere \cite{Ackermann2012,Aguilar2014}, 
pulsars \cite{Hooper2009,Profumo2012}, and microquasars \cite{Siegert2016}. 
Moreover, e-p plasmas can be created by
ultra-intense laser interaction with matter in the laboratory
\cite{Liang1998,Bell2008,Chen2009,Sarri2013,Sarri2015}. The long-lived runaway
positrons can also be generated in post-disruption tokamak plasmas
\cite{Helander2003}. In dense astrophysical environments, ions usually exist
in addition to electrons and positrons, for example, nearby hot white dwarfs
and microquasar \cite{Kotani1996,Lallement2011}. Energetic electrons, accelerated to high suprathermal energies, are also found to be produced in ultra-intense laser fields \cite{Liang1998},
tokamaks \cite{Savrukhin2001}, the solar wind \cite{Holman2003,Fisk2006}, and the
Earth's magnetosphere \cite{Vasyliunas1968}. In particular, the energy distribution of suprathermal electrons in solar flares was found to be well described by a power law with a maximum high-energy cutoff of 3 GeV \cite{Fisk2006}. Hence, studying e-p pair plasmas with suprathermal electrons are important for both laboratory and astrophysical plasmas.

Electrostatic waves usually occur in a plasma containing distinct electron
populations with different temperatures
\cite{Watanabe1977,Tokar1984,Gary1985,Mace1990}, namely \textit{cool} electrons,
$T_{c}$, and \textit{hot} electrons, $T_{h}$. The cool electron motion provides the
inertia required to maintain electrostatic oscillations, while the hot
electron pressure produces the restoring force for 
electrostatic waves propagating at a phase speed between the cool and hot
electron thermal velocities. In such a plasma, the ions can be assumed to make
a stationary background providing charge neutrality. It is found that Landau
damping is minimized if the cool electron fraction of the total number density of
electrons is in the range of $0.2\lesssim n_{c}/(n_{c}+n_{h})\lesssim0.8$ and
the hot electron temperature is much higher than the cool electron temperature
$T_{h}/T_{c} \gg10$~~\cite{Gary1985,Mace1990,Mace1999,Baluku2011}. The
dynamics of electron-acoustic waves in a two-electron-temperature plasma have
been studied by many authors
\cite{Watanabe1977,Tokar1984,Gary1985,Mace1990,Mace1999}. Moreover, linear
and nonlinear studies of electron-acoustic waves in the presence of
suprathermal (or non-thermal) electrons have received a great deal of interest in recent years,
both in unmagnetized \cite{Baluku2011,Sultana2011,Danehkar2011}
and in magnetized plasmas \cite{Sultana2010,Sultana2012}. Negative polarity
electrostatic wave structures were found to exist in a two-electron-temperature
plasma with excess suprathermal electrons \cite{Sultana2010,Danehkar2011},
which are associated with the inertia of mobile cold electrons. However,
positive polarity electrostatic waves moving at velocities comparable to electrons
have been reported in the auroral magnetosphere
\cite{Muschietti1999,Cattell2002}. Inclusion of a beam component
\cite{Berthomier2000,Mace2001} or finite
inertia \cite{Cattaert2005,Verheest2005} may lead to a positive polarity electrostatic
wave. Alternatively, a fraction of mobile positrons (or electron holes), which are
created by the solar wind, may maintain the inertia for the propagation of
positive polarity electrostatic waves. Interestingly, a considerable fraction of positrons 
has been recently observed at the solar wind: $\phi(\mathrm{e}^{+})/(\phi(\mathrm{e}^{+})+\phi(\mathrm{e}^{-}))\lesssim 0.1$ at energies 0.04--1\,GeV \cite{Clem1996,Protheroe1982,Moskalenko1998,Clem2000,Adriani2009,Adriani2011}. Moreover, a significant
positron density has been measured in laser-plasma experiences: $n(\mathrm{e}^{+})/(n(\mathrm{e}^{+})+n(\mathrm{e}^{-}))\sim0.05$--0.1 in $\sim10$\,MeV \cite{Chen2009} and $\sim0.01$--0.1 in $150$\,MeV \cite{Sarri2013}. The experimental temperature of positrons was measured to be roughly half of the effective electron temperature in ultra-intense laser fields \cite{Chen2009} 
(In this paper, $T_{p} \sim T_{c}$, while $T_{p}/T_{h} \ll0.1$ to minimize Landau damping; see Ref.~\onlinecite{Verheest1996}). The positron effect may have important 
implications for the dynamics of positive polarity electrostatic waves of the auroral magnetosphere. 

The propagation of electrostatic waves in e-p pair plasmas can also be supported by the inertia of 
mobile cool positrons, while background hot electrons act as the restoring force. 
Interestingly, the coherent microwave radiation has been reported in pulsars \cite{Sturrock1971,Ruderman1975}, which is assumed to be originated from electric fields of e-p pair plasma over the polar caps of neutron stars \cite{Arons1979,Arons1983}. 
It was proposed that coherent pulsar radio emissions could be due to nonlinear electrostatic solitary oscillations generated by effective electron-positron streams on the polar caps in rotating magnetized neutron stars \cite{Melikidze2000}.
Studies of e-p pair plasmas demonstrated that electrostatic solitary waves can be generated 
\cite{Verheest1996}, though a Maxwellian distribution was assumed.  
A number of papers have also been devoted to the linear and nonlinear dynamics of 
electron-acoustic waves \cite{Jilani2015,Rafat2015}, electrostatic waves \cite{Jao2012,Lazarus2012,Saberian2013,Jao2014,Jilani2014,Mugemana2014,Saberian2014,Chatterjee2015}, and in the presence of suprathermal (and non-thermal) electrons
\cite{Saberian2013,Jilani2015,Jilani2014} in e-p pair plasmas.  
Moreover, the propagation of ion-acoustic waves 
\citep{Esfandyari-Kalejahi2009,Chawla2010,Baluku2011a,Ferdousi2015},
and dust-acoustic waves \citep{Jehan2009,Esfandyari-Kalejahi2012,Saberian2017} have recently been studied in e-p plasmas. 
However, the nonlinear dynamics and the existence domains of electrostatic solitary waves  have not fully been investigated in the presence of positrons. 
It is important to study the occurrence of electrostatic solitary wave structures in e-p pair plasmas with suprathermal electrons, which may lead
to the (co-)existence of positive and negative polarity electrostatic waves similar to what
observed in the Earth's magnetosphere \cite{Muschietti1999,Cattell2002}, as well as a possible explanation for coherent pulsar radio emissions \cite{Melikidze2000}. 

In this paper, we aim to explore the effect of mobile cool positrons (electron holes)
on electrostatic solitary waves in an e-p pair plasma with suprathermal electrons. In Section~\ref{sec:model}, 
a two-fluid model is presented. In Section \ref{sec:DR}, a dispersion relation
is derived. In Section \ref{sec:nonlinear}, a nonlinear pseudopotential (Sagdeev)
method is used to investigate the existence of large-amplitude electrostatic solitary
waves. Section \ref{sec:investigation} is devoted to a parametric
investigation of the nonlinear form and the characteristics of electrostatic
solitary wave structures.
Finally, our results are summarized in the concluding section
\ref{sec:conclusion}.

\section{Theoretical Model\label{sec:model}}

We consider a 1-D collisionless, four-component
plasma consisting of cool inertial background electrons (at temperature
$T_{c}\neq0$), mobile cool positrons (or electron holes; at temperature $T_{p}\neq0$),
inertialess hot suprathermal electrons modeled by a $\kappa$-distribution (at
temperature $T_{h}\gg T_{c},T_{p}$), and uniformly distributed stationary ions.

The cool electrons and positrons are governed by the following fluid
equations:
\begin{align}
&  \frac{\partial n_{c}}{\partial t}+\frac{\partial(n_{c}u_{c})}{\partial
x}=0,\label{eq_1}\\
&  \frac{\partial u_{c}}{\partial t}+u_{c}\frac{\partial u_{c}}{\partial
x}=\frac{e}{m_{e}}\frac{\partial\phi}{\partial x}-\frac{1}{m_{e}n_{c}}
\frac{\partial p_{c}}{\partial x},\label{eq_2}\\
&  \frac{\partial p_{c}}{\partial t}+u_{c}\frac{\partial p_{c}}{\partial
x}+\gamma p_{c}\frac{\partial u_{c}}{\partial x}=0,\label{eq_3}\\
&  \frac{\partial n_{p}}{\partial t}+\frac{\partial(n_{p}u_{p})}{\partial
x}=0,\label{eq_4}\\
&  \frac{\partial u_{p}}{\partial t}+u_{p}\frac{\partial u_{p}}{\partial
x}=-\frac{e}{m_{p}}\frac{\partial\phi}{\partial x}-\frac{1}{m_{p}n_{p}}
\frac{\partial p_{p}}{\partial x},\label{eq_5}\\
&  \frac{\partial p_{p}}{\partial t}+u_{p}\frac{\partial p_{p}}{\partial
x}+\gamma p_{p}\frac{\partial u_{p}}{\partial x}=0, \label{eq_6}
\end{align}
where $n$, $u$ and $p$ are the number density, the velocity and the pressure
of the cool electrons and positrons (denoted by indices `\textit{c}' and
`\textit{p}', respectively), $\phi$ is the electrostatic wave potential, $e$
the elementary charge, $m_{e}$ the electron mass, $m_{p}$ the positron mass,
and $\gamma=(f+2)/f$ denotes the specific heat ratio for $f$ degrees of
freedom. For the adiabatic cool electrons and positrons in one-dimensional ($f=1$), we get
$\gamma=3$. Through this paper, we assume that $m_{e}=m_{p}$.

Following Eq.~1 in Ref.~\onlinecite{Danehkar2011}, the $\kappa$-distribution
expression is obtained for the number density of the hot suprathermal
electrons:
\begin{equation}
n_{h}(\phi)=n_{h,0}\left[  1-\frac{e\phi}{k_{B}T_{h}(\kappa-\tfrac{3}{2}
)}\right]  ^{-\kappa+1/2}\,, \label{eq_7}
\end{equation}
where $n_{h,0}$ and $T_{h}$ are the equilibrium number density and the
temperature of the hot electrons, respectively, $k_{B}$ the Boltzmann
constant, and the spectral index $\kappa$ measures the deviation from thermal equilibrium.
For reality of the characteristic modified thermal velocity, $\left[  (2\kappa-3)k_{B}T_{h}/\kappa m_{e}\right]  ^{1/2}$, the spectral index must
take $\kappa>3/2$. The suprathermality is measured by the spectral index $\kappa$, 
describing how it deviates from a Maxwellian distribution, i.e., low values of $\kappa$ 
are associated with a significant suprathermality; on the other hand, a 
Maxwellian distribution is recovered in the limit $\kappa\rightarrow\infty$. 

The ions are assumed to be immobile in a uniform state, i.e., $n_{i}=n_{i,0}=$
const. at all times, where $n_{i,0}$ is the undisturbed ion density. The
plasma is quasi-neutral at equilibrium, so $Zn_{i,0}+n_{p,0}=n_{c,0}+n_{h,0}$,
that implies
\begin{equation}
Z{n_{i,0}}/{n_{c,0}}=1+\alpha-\beta, \label{eq_8}
\end{equation}
where we have defined the hot-to-cool electron density ratio as $\alpha
=n_{h,0}/n_{c,0}$, and the positron-to-cool electron density ratio as
$\beta=n_{p,0}/n_{c,0}$, while $n_{c,0}$ and $n_{p,0}$ are the equilibrium
number densities of the cool electrons and positrons, respectively.
Electrostatic waves are weakly damped in the range of $0.2\lesssim
n_{c,0}/(n_{c,0}+n_{h,0})\lesssim0.8$
\cite{Gary1985,Mace1990,Mace1999,Baluku2011}, so $0.25\leqslant
\alpha\leqslant4$. This region may permit the propagation of nonlinear
electrostatic structures. As the positron fraction has been measured to be $\phi(\mathrm{e}^{+})/(\phi(\mathrm{e}^{+}
)+\phi(\mathrm{e}^{-}))\lesssim0.1$  in low energy solar wind observations \cite{Clem1996,Protheroe1982,Moskalenko1998,Adriani2009} and $n(\mathrm{e}^{+})/(n(\mathrm{e}^{+})+n(\mathrm{e}^{-}))\sim0.05$--0.1 in some laser-plasma experiences \cite{Chen2009,Sarri2013}, we assume $\beta\lesssim0.06$.

All four components are coupled via the Poisson's equation:
\begin{equation}
\frac{\partial^{2}\phi}{\partial x^{2}}=-\frac{e}{\varepsilon_{0}}\left(
Zn_{i}+n_{p}-n_{c}-n_{h}\right)  , \label{eq_10}
\end{equation}
where $\varepsilon_{0}$ is the permittivity constant.

Scaling by appropriate quantities, we arrive at a fluid system of our model in
a dimensionless form for the cool electrons and the positrons, respectively:
\begin{align}
&  \frac{\partial n}{\partial t}+\frac{\partial(nu)}{\partial x}
=0,\label{eq_11}\\
&  \frac{\partial u}{\partial t}+u\frac{\partial u}{\partial x}=\frac
{\partial\phi}{\partial x}-\frac{\sigma}{n}\frac{\partial p}{\partial
x},\label{eq_12}\\
&  \frac{\partial p}{\partial t}+u\frac{\partial p}{\partial x}+3p\frac
{\partial u}{\partial x}=0, \label{eq_13}\\
&  \frac{\partial n_{p}}{\partial t}+\frac{\partial(n_{p}u_{p})}{\partial
x}=0,\label{eq_14}\\
&  \frac{\partial u_{p}}{\partial t}+u_{p}\frac{\partial u_{p}}{\partial
x}=-\frac{\partial\phi}{\partial x}-\frac{\theta}{n_{p}}\frac{\partial p_{p}
}{\partial x},\label{eq_15}\\
&  \frac{\partial p_{p}}{\partial t}+u_{p}\frac{\partial p_{p}}{\partial
x}+3p_{p}\frac{\partial u_{p}}{\partial x}=0. \label{eq_16}
\end{align}
The dimensionless Poisson's equation takes the following form:
\begin{align}
\frac{\partial^{2}\phi}{\partial x^{2}} =  & -\left(  1+\alpha-\beta\right) +n-\beta n_{p} \nonumber\\
 &  +\alpha\left(  1-\frac{\phi}{\kappa-\tfrac{3}{2}}\right)^{-\kappa+1/2}, \label{eq_17} 
\end{align}
where $n$ and $n_{p}$ denote the fluid density variables of the cool electrons
and positrons normalized with respect to $n_{c,0}$ and $n_{p,0}$,
respectively, $u$ and $u_{p}$ the velocity variables of the cool electrons and
positrons scaled by the hot electron thermal speed $c_{th}=\left(  k_{B}
T_{h}/m_{e}\right)  ^{1/2}$, $p$ and $p_{p}$ the pressure variables of the
cool electrons and positrons normalized with respect to $n_{c,0}k_{B}T_{c}$
and $n_{p,0}k_{B}T_{p}$, respectively, and the wave potential $\phi$ by $k_{B}
T_{h}/e$, time and space scaled by the plasma period $\omega_{pc}^{-1}=\left(
n_{c,0}e^{2}/\varepsilon_{0}m_{e}\right)  ^{-1/2}$ and the characteristic
length $\lambda_{0}=\left(  \varepsilon_{0}k_{B}T_{h}/n_{c,0}e^{2}\right)
^{1/2}$, respectively. We have defined the cool-to-hot electron temperature
ratio as $\sigma=T_{c}/T_{h}$, and the positron-to-hot electron temperature
ratio as $\theta=T_{p}/T_{h}$. Landau damping is minimized if $\sigma=T_{c}/T_{h}\ll0.1$
\cite{Gary1985,Mace1990,Mace1999}, and the same for the cool positrons, i.e.,
$\theta=T_{p}/T_{h}\ll0.1$ (see Ref.~\onlinecite{Verheest1996}), and typically $T_{p}/T_{c} \sim0.5$ in some laser-plasma experiences \cite{Chen2009}. 

\section{Linear Dispersion Relation\label{sec:DR}}

To obtained the linear dispersion relation, we substitute linearized forms of
Eqs.~(\ref{eq_11})--(\ref{eq_16}) to the Poisson's equation (\ref{eq_17}) and restrict up to the first order, which yield:
\begin{equation}
1+\frac{k_{D,\kappa}^{2}}{k^{2}}=\frac{1}{\omega^{2}-3{\sigma}k^{2}}  +\frac{\beta}{\omega^{2}-3\theta k^{2}}  . \label{eq_23_1}
\end{equation}
where the appearance of a normalized $\kappa$-dependent screening factor
(scaled Debye wavenumber) $k_{D,\kappa}$ is defined by
\begin{equation}
k_{D,\kappa}\equiv\dfrac{1}{\lambda_{D,\kappa}}\equiv\left[  \dfrac
{\alpha(\kappa-\frac{1}{2})}{\kappa-\tfrac{3}{2}}\right]  ^{1/2}\,.
\label{eq_24}
\end{equation}
Eq.~(\ref{eq_23_1}) contains a suprathermality term, a Langmuir wave mode, and a positron wave mode.
In this equation, $\sqrt{3\sigma}$ corresponds to the normalized cool electron thermal velocity, and $\sqrt{3\theta}$ is associated with the normalized positron thermal velocity. In the absence of the hot electrons, a linear dispersion is derived in agreement with Eq. 4 im Ref.~\onlinecite{Verheest2006} and Eq. 8 im Ref.~\onlinecite{Jao2012}. It is seen that the phase speed increases
with higher cool-to-hot electron temperature ratio $\sigma=T_{c}/T_{h}$, in
agreement with what found previously \cite{Danehkar2011}. In the limit
$\beta\rightarrow0$ (in the absence of the positrons), we obtain
Eq.~(14) of \onlinecite{Danehkar2011}. From Eq.~(\ref{eq_23_1}), we
see that the frequency $\omega(k)$, and hence the phase speed, increases
with higher positron-to-hot electron temperature ratio $\theta=T_{p}/T_{h}$.
However, this linear thermal effect may not be noticeable due to the range of
parameters adopted here, i.e., low values of the positron-to-cool electron
density ratio ($\beta\lesssim0.06$).

\begin{figure}
\begin{center}
\includegraphics[width=3.0in]{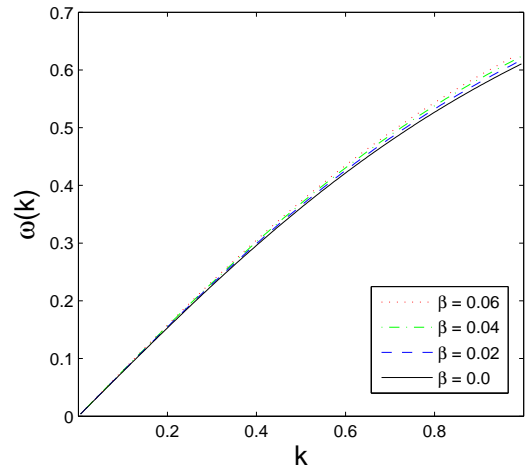}
\caption{Variation of the dispersion curve for different values
of the positron-to-cool electron density ratio $\beta$. 
Curves from bottom to top: $\beta=0.0$ (solid), $0.02$ (dashed),
$0.04$ (dot-dashed curve), and $0.06$ (dotted curve). Here, $\alpha=1$,
$\kappa=3$, $\theta=0.01$ and $\sigma=0$.}
\label{fig1}
\end{center}
\end{figure}

In Figure \ref{fig1}, we plot the dispersion curve (\ref{eq_23_1}) in the
electron cold limit ($\sigma=0$), showing the effect of varying the values of
the positron-to-cool electron charge density ratio $\beta$. It can be seen that an
increase in the positron parameter $\beta$ weakly increases the phase speed
($\omega/k$). Noting that the positron fraction in astrophysical plasmas, for
example in the solar wind \cite{Adriani2009}, is very small, its effect on the small-amplitude wave solutions is
negligible. Otherwise, a large fraction of the positron fraction can
significantly modify the dispersion relation. Previously, we found that the
dispersion relation and the condition for Landau damping are considerably
changed, when the plasma is dominated by hot $\kappa$-distributed electrons (see Figure 1 in Ref.~\onlinecite{Danehkar2011}). Similarly,
increasing the number density of suprathermal hot electrons or/and the
suprathermality (decreasing $\kappa$) also decreases the phase speed.

\section{Nonlinear Analysis\label{sec:nonlinear}}

To obtain nonlinear wave solutions, we consider all fluid variables in a
stationary frame traveling at a constant normalized velocity $M$ (to be
referred to as the Mach number), which implies the transformation $\xi=x-Mt$.
This replaces the space and time derivatives with $\partial/\partial x=d/d\xi$
and $\partial/\partial t=-Md/d\xi$, respectively. Now equations (\ref{eq_11})
to (\ref{eq_17}) take the following form:
\begin{align}
&  -M\dfrac{dn}{d\xi}+\frac{d(nu)}{d\xi}=0,\label{eq_26}\\
&  -M\dfrac{du}{d\xi}+u\dfrac{du}{d\xi}=\dfrac{d\phi}{d\xi}-\frac{\sigma}
{n}\dfrac{dp}{d\xi},\label{eq_27}\\
&  -M\dfrac{dp}{d\xi}+u\dfrac{dp}{d\xi}+3p\dfrac{du}{d\xi}=0,\label{eq_28}\\
&  -M\dfrac{dn_{p}}{d\xi}+\frac{d(n_{p}u_{p})}{d\xi}=0,\label{eq_29}\\
&  -M\dfrac{du_{p}}{d\xi}+u_{p}\dfrac{du_{p}}{d\xi}=-\dfrac{d\phi}{d\xi}
-\frac{\theta}{n_{p}}\dfrac{dp_{p}}{d\xi},\label{eq_30}\\
&  -M\dfrac{dp_{p}}{d\xi}+u_{p}\dfrac{dp_{p}}{d\xi}+3p_{p}\dfrac{du_{p}}{d\xi
}=0, \label{eq_31}\\
\dfrac{d^{2}\phi}{d\xi^{2}}=  &  -\left(  1+\alpha-\beta\right)  +n-\beta
n_{p}\nonumber\\
&  +\alpha\left(  1-\frac{\phi}{\kappa-\tfrac{3}{2}}\right)  ^{-\kappa+1/2},
\label{eq_32}
\end{align}
The equilibrium state is assumed to be reached at both infinities
($\xi\rightarrow\pm\infty$). Accordingly, we integrate Eqs.~(\ref{eq_26})--(\ref{eq_31}), apply the boundary conditions $n=1$, $p=1$, $u=0$, $n_{p}=1$, $p_{p}=1 $, $u_{b}=0$ and $\phi=0$ at infinities, and obtain
\begin{align}
&  u=M[1-(1/n)],\label{eq_33}\\
&  u={M-(M}^{2}{+2\phi-3n^{2}\sigma+3\sigma})^{1/2},\label{eq_34}\\
&  u_{p}=M[1-(1/n_{p})],\label{eq_35}\\
&  u_{p}=M-(M^{2}-2\phi-3n_{p}^{2}\theta+3\theta)^{1/2},\label{eq_36}\\
&
\begin{array}
[c]{cc}%
p=n^{3},\text{ \ } & p_{p}=n_{p}^{3}.
\end{array}
\label{eq_37}
\end{align}
Combining Eqs.~(\ref{eq_33})--(\ref{eq_37}), one obtains the following
biquadratic equations for the cool electron density and the positron density,
respectively,
\begin{align}
& 3\sigma n^{4}-({M}^{2}{+2\phi+3\sigma)n^{2}}+{M}^{2}=0,  \label{eq_38} \\ 
& 3\theta n_{p}^{4}-(M^{2}-2\phi+3\theta)n_{p}^{2}+M^{2}=0. \label{eq_39}
\end{align}
Eqs.~(\ref{eq_38}) and ~(\ref{eq_39}) are respectively solved as follows:
\begin{align}
n  &  =\frac{1}{2\sqrt{3\sigma}}\left[  {2\phi+}({M+\sqrt{3{\sigma}}}
)^{2}\right]  ^{1/2}\nonumber\\
&  \pm\frac{1}{2\sqrt{3\sigma}}{\left[  {2\phi+({M-\sqrt{3{\sigma}}}}
)^{2}\right]  }^{1/2}, \label{eq_40}\\
{n}_{p}  &  {=}\frac{1}{2\sqrt{3\theta}}\left[  -{2\phi}+(M+\sqrt{3\theta
})^{2}\right]  ^{1/2}\nonumber\\
&  \pm\frac{1}{2\sqrt{3\theta}}\left[  -{2\phi}+(M-\sqrt{3\theta})^{2}\right]
^{1/2}. \label{eq_41}
\end{align}
Eq.~(\ref{eq_40}) agrees with Eq. (29) derived in Ref.~\onlinecite{Danehkar2011}. 
From the boundary conditions, $n_{c}=n_{p}=1$ at $\phi=0$, it follows that the
negative sign must be taken in equations (\ref{eq_40}) and (\ref{eq_41}).
Moreover, the cool electrons and positrons are assumed to be supersonic for
$M>\sqrt{3\sigma}$ and $M>\sqrt{3\theta}$, respectively, while the hot
electrons are subsonic for $M<1$.

Reality of the cool electron density variable imposes the requirement 
${2\phi+({M-}\sqrt{3{\sigma}})^{2}>0}$ that implies a lower boundary on the
electrostatic potential value
$\phi>\phi_{{\rm max}(-)}=-\frac{1}{2}({M-}\sqrt{3{\sigma}})^{2}$ associated with
negative polarity solitary structures. However, reality of the positron density
variable imposes ${-2\phi}+(M-\sqrt{3\theta})^{2}{>0}$, implying a higher
boundary on the electrostatic potential value
$\phi< \phi_{{\rm max}(+)}=\frac{1}{2}({M-}\sqrt{3\theta})^{2}$ associated with
positive polarity solitary structures.

Substituting Eqs.~(\ref{eq_40})--(\ref{eq_41}) into the Poisson's equation
(\ref{eq_17}), multiplying the resulting equation by $d\phi/d\xi$, integrating
and taking into account the conditions at infinities ($d\phi/d\xi\rightarrow
0$) yield a pseudo-energy balance equation:
\begin{equation}
\frac{1}{2}\left(  \frac{d\phi}{d\xi}\right)  ^{2}+\Psi(\phi)=0, \label{eq_42}
\end{equation}
where the Sagdeev pseudopotential $\Psi(\phi)$ is given by
\begin{align}
\Psi(\phi)=  &  \alpha\left[  1-\left(  1+\frac{\phi}{-\kappa+\tfrac{3}{2}
}\right)  ^{-\kappa+3/2}\right]  +(1+\alpha-\beta)\phi\nonumber\\
&  +\frac{1}{6\sqrt{3{\sigma}}}\left[  ({M+}\sqrt{3{\sigma}})^{3}-{{({M-}
\sqrt{3{\sigma}})^{3}}}\right. \nonumber\\
&  \left.  -({2\phi+}[{M+}\sqrt{3{\sigma}}]^{2})^{3/2}\right. \nonumber\\
&  \left.  +{({2\phi+[{M-}\sqrt{3{\sigma}}]^{2}}})^{3/2}\right] \nonumber\\
&  -\frac{\beta}{6\sqrt{3\theta}}\left[  (M_{\text{ }}{+}\sqrt{3\theta}
)^{3}-{{(M{-}\sqrt{3\theta})^{3}}}\right. \nonumber\\
&  \left.  -(-{2\phi+}[M{+}\sqrt{3\theta}]^{2})^{3/2}\right. \nonumber\\
&  \left.  +{(-{2\phi+[M{-}\sqrt{3\theta}]^{2}}})^{3/2}\right]  .
\label{eq_43}
\end{align}
In the absence of the positrons ($\beta\rightarrow0$), we exactly recover
the pseudopotential equation derived for electron-acoustic waves with suprathermal electrons \cite{Danehkar2011}.

For the existence of solitons, we require that the origin at $\phi=0$ is a
root and a local maximum of $\Psi$ in Eq.~(\ref{eq_43}), i.e., $\Psi(\phi)=0$,
$\Psi^{\prime}(\phi)=0$ and $\Psi^{\prime\prime}(\phi)<0$ at $\phi=0$, where
primes denote derivatives with respect to $\phi$. It is easily seen that the
first two constraints are satisfied. We thus impose the condition
$F_{1}(M)=-\Psi^{\prime\prime}(\phi)|_{\phi=0}>0$, and we get
\begin{equation}
F_{1}(M)=\frac{\alpha(\kappa-\frac{1}{2})}{\kappa-\tfrac{3}{2}}-\frac
{1}{(M^{2}-3\sigma)}-\frac{\beta}{(M^{2}-3\theta)}. \label{eq_44}
\end{equation}
Eq.~(\ref{eq_44}) provides the minimum value for the Mach number,
$M_{1}(\kappa,\alpha,\sigma,\beta,\theta)$. In the limit $\beta
\rightarrow0$ (without the positrons), equation (\ref{eq_44}) takes the form of
Eq.~(34) in Ref.~\onlinecite{Danehkar2011}.

An upper limit for $M$ is determined from the fact that the cool electron
density becomes complex at negative potentials lower than $\phi_{{\rm max}(-)}=-\frac{1}{2}(  {M-}
\sqrt{3{\sigma}})  ^{2}$ for negative polarity waves, and the cool 
positron density at positive potentials higher than  $\phi
_{{\rm max}(+)}=\frac{1}{2}(  M{-}\sqrt{3{\theta}})  ^{2}$ for positive
polarity waves. Thus, the largest negative soliton amplitude satisfies
$F_{2}(M)=\Psi(\phi)|_{\phi=\phi_{{\rm max}(-)}}>0$, whereas the largest positive
soliton amplitude fulfills $F_{2}(M)=\Psi(\phi)|_{\phi=\phi_{{\rm max}(+)}}>0$.
These yield the following equation for the upper limit in $M$ for negative
polarity electrostatic soliton existence associated with cool electrons,
\begin{align}
F_{2}^{(-)}(M)  &  =-\tfrac{1}{2}(1+\alpha-\beta)({M-}\sqrt{3{\sigma}}
)^{2}+M^{2}+\sigma\nonumber\\
&  +\alpha\left[  1-\left(  1+\frac{[{M-}\sqrt{3{\sigma}}]^{2}}{2\kappa
-3}\right)  ^{-\kappa+3/2}\right] \nonumber\\
&  -\frac{\beta}{6\sqrt{3\theta}}\left(  {[}({{M-\sqrt{3\theta})^{2}}-}
({{{M-}\sqrt{3{\sigma}})^{2}}}]^{3/2}\right. \nonumber\\
&  \left.  -[(M{+}\sqrt{3\theta})^{2}-{({M-}\sqrt{3{\sigma}}){^{2}}}
]^{3/2}\right) \nonumber\\
&  -\beta M^{2}{-}\beta\theta-\tfrac{4}{3}M^{3/2}\left(  3\sigma\right)
^{1/4}, \label{eq_45}
\end{align}
and the following equation for positive polarity electrostatic soliton existence associated with positrons,
\begin{align}
F_{2}^{(+)}(M)  &  =\tfrac{1}{2}(1+\alpha-\beta)(M{-}\sqrt{3{\theta}}
)^{2}+M^{2}+\sigma\nonumber\\
&  +\alpha\left(  1-\left[  1-\frac{(M{-}\sqrt{3{\theta}})^{2}}{2\kappa
-3}\right]  ^{-\kappa+3/2}\right) \nonumber\\
&  +\frac{1}{6\sqrt{3{\sigma}}}\left(  {[({{{M-}\sqrt{3{\sigma}})^{2}}+}
(M{-}\sqrt{3{\theta}}){^{2}}}]^{3/2}\right. \nonumber\\
&  \left.  -[({M+}\sqrt{3{\sigma}})^{2}+(M{-}\sqrt{3{\theta}})^{2}
]^{3/2}\right) \nonumber\\
&  -\beta M^{2}-\beta\theta+\tfrac{4}{3}\beta M^{3/2}\left(  3\theta\right)
^{1/4}. \label{eq_46}
\end{align}
Solving equations (\ref{eq_45}) and (\ref{eq_46}) provide the upper limit
$M_{2}(\kappa,\alpha,\sigma,\beta,\theta)$ for acceptable values of the Mach
number for negative and positive polarity solitons to exist. The cool electrons can generally support a
negative supersonic electrostatic wave, while the positrons may provide the inertia to support a
positive polarity electrostatic wave. Hence, the upper limit of negative polarity electrostatic
solitons can be determined from Eq.~(\ref{eq_45}), while the upper limit of
positive polarity electrostatic solitons may be obtained from Eq.~(\ref{eq_46}). In the
absence of the positrons, Eq.~(\ref{eq_45}) yields exactly Eq.~(36) in Ref.~\onlinecite{Danehkar2011}. 
Taking a Maxwellian distribution ($\kappa\rightarrow\infty$) and without the positrons ($\beta\rightarrow0$), equations~(\ref{eq_44}) and (\ref{eq_45}) take the form of Eqs.~(37) and (38) in Ref.~\onlinecite{Danehkar2011}.

\begin{figure}
\begin{center}
\includegraphics[width=3.0in]{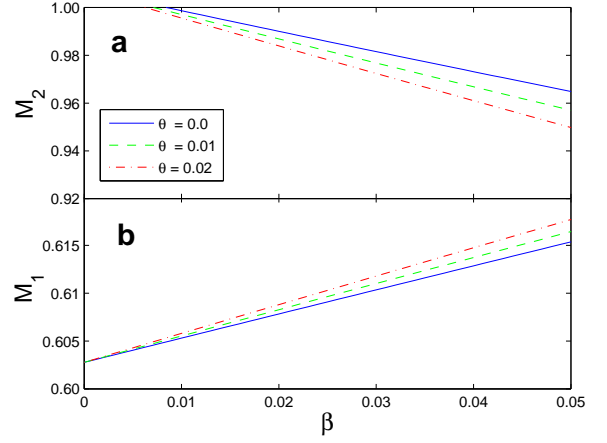}
\caption{Variation of the lower limit $M_{1}$ (lower curves;
panel b) and the upper limit $M_{2}$ (upper curves; panel a) of the negative
polarity electrostatic solitons with the positron-to-cool electron density ratio $\beta$ for different values of the positron-to-hot electron temperature ratio $\theta$. Solitons may exist
for values of the Mach number $M$ in the region between the lower and the
upper curve(s) of the same style/color. (a-b) Curves: $\theta=0.0$\ (solid),
$0.01$\ (dashed), and $0.02$\ (dot-dashed). Here, $\kappa=2$, $\alpha=1$ and
$\sigma=0.01$.}
\label{fig2}
\end{center}
\end{figure}

Figure \ref{fig2} shows the range of allowed Mach numbers for negative
polarity electrostatic solitary waves with different parameters: the positron-to-hot
electron temperature ratio, $\theta$, and the positron-to-cool electron
density ratio, $\beta$. The lower limit ($M_{1}$) and the upper limit ($M_{2}
$) of Mach numbers are obtained from numerically solving equations
(\ref{eq_44}) and (\ref{eq_45}), respectively. We see that there is a small
difference between the model including the positrons and the model without the
positrons ($\beta\rightarrow0$). As the positron is assumed to have a
very small fraction of the total charge ($\beta\lesssim0.06$) and a cool
temperature ($\theta\ll0.1$), they cannot have a significant role in the
dynamics of electron-acoustic waves in the model adopted here. Hence, the
existence domain of electron-acoustic (negative polarity electrostatic) solitary waves are not largely affected by the
cool positrons.

\begin{figure}
\begin{center}
\includegraphics[width=3.0in]{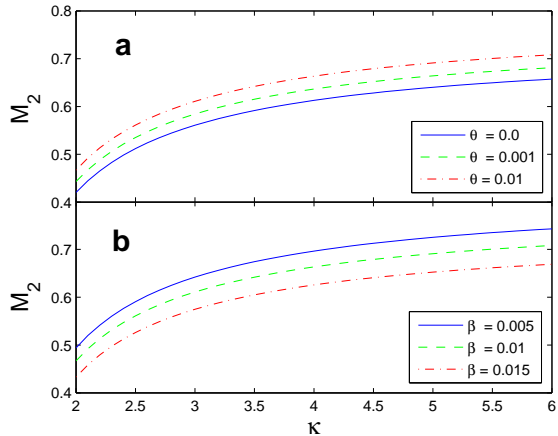}
\caption{Variation of the upper limit $M_{2}$ of the positive
polarity electrostatic solitons with the spectral index $\kappa$ for different values of (a) the positron-to-hot electron temperature ratio $\theta$ and (b)
the positron-to-cool electron density ratio $\beta$.  Upper panel: $\theta=0.0$\ (solid), $0.001$\ (dashed), and
$0.01$\ (dot-dashed). Here, $\alpha=1$ and $\beta=\sigma=0.01$. Lower panel:
$\beta=0.005$\ (solid), $0.01$\ (dashed), and $0.015$\ (dot-dashed). Here,
$\alpha=1$ and $\sigma=\theta=0.01$.}
\label{fig4}
\end{center}
\end{figure}

The soliton existence regions for positive polarity electrostatic solitary waves are
shown in Fig.~\ref{fig4} for different parameters. Solitary
structures of the electrostatic potential may occur in the range
$M_{1}<M<M_{2}$, which depends on the parameters $\theta$, $\beta$, and
$\kappa$. Moreover, we assume that the cool electrons and positrons are
supersonic (${M>}\sqrt{3{\sigma}}$ and ${M>}\sqrt{3{\theta}}$, respectively),
while the hot electrons are subsonic (${M<1}$). We used Eq.~(\ref{eq_44}) to
obtain the lower limit for negative polarity solitons. This equation may also have another
solution, which could yield the lower Mach number limit for positive polarity solitary
structures. However, we noticed that Mach numbers of positive polarity solitons cannot
be constrained by Eq.~(\ref{eq_44}) due to the small values of the density ratio 
$\beta$. Therefore, the lower limit ($M_{1}$) is found to be at about 
$\sqrt{3{\sigma}}$. The positive potential solitons numerically derived from
Eq.~(\ref{eq_43}) cannot also produce any solutions for Mach numbers less than
$\sqrt{3{\sigma}}$ in the adopted parameter ranges of the positrons.

As seen in Fig.~\ref{fig4}, the upper limit ($M_{2}$) of positive polarity solitons 
is slightly increased with an increase in the positron-to-hot electron temperature ratio
$\theta$ and a decrease in the positron-to-cool electron density
ratio $\beta$. However, the effect is not significant, and also dissimilar to how 
the hot-to-cool electron density ratio ($\alpha$) affects electron-acoustic waves \cite{Danehkar2011}. 
This negligible effect is mostly attributed to the
small fraction of positrons and their cool temperatures in the e-p plasma system.

Figure \ref{fig4} also depicts the upper limit ($M_{2}$) of allowed Mach numbers
as a function of $\kappa$, for various values of $\theta$ and $\beta$. As seen, increasing $\kappa$ toward a Maxwellian distribution
($\kappa\rightarrow\infty$) increases the upper limit ($M_{2}$) and broadens
the Mach number range. It can be seen that positive polarity solitons are generated in
narrower ranges of Mach numbers as hot electron suprathermality becomes
stronger. This conclusion is similar to what found in electron-acoustic solitary waves with suprathermal electrons
\cite{Danehkar2011}.

\section{Nonlinear Wave Structures \label{sec:investigation}}

To consider the nonlinear features of electrostatic wave structures, we
have numerically solved Eq.~(\ref{eq_43}) for various plasma parameters, in
order to investigate their effects. We found that both negative and positive
electric potentials arise in the ranges of allowed Mach numbers obtained for
negative and positive polarity soliton existence domains in Section \ref{sec:nonlinear}.

\begin{figure}
\begin{center}
\includegraphics[height=5.4in]{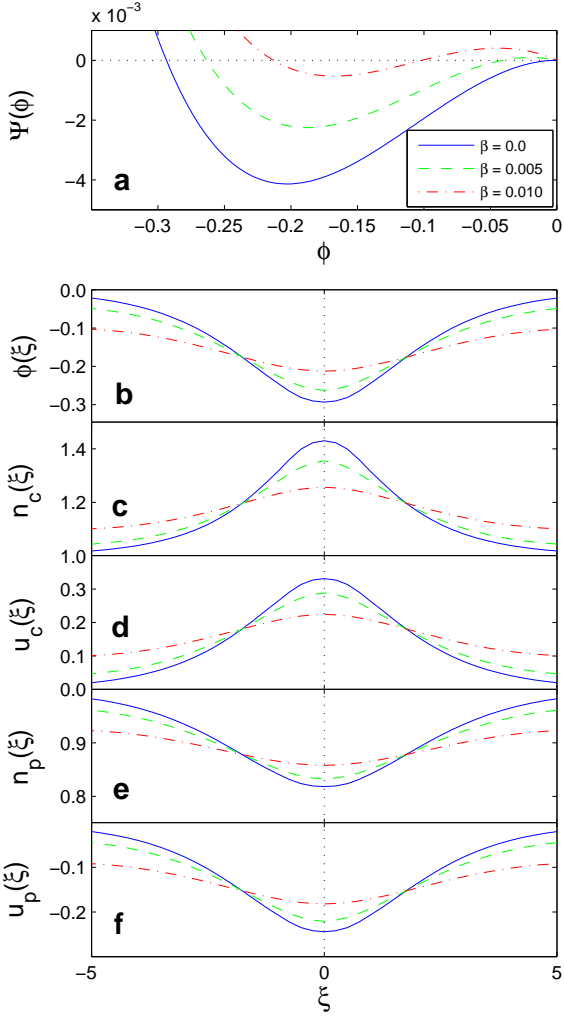}
\caption{(a) The pseudopotential $\Psi(\phi)$ of negative
polarity electrostatic solitons and the associated solutions: (b) electric potential
pulse $\phi$, (c) density $n_{c}$ and (d) velocity $u_{c}$ of the cool electron fluid,
and (e) density $n_{p}$ and (f) velocity $u_{p}$ of the positron fluid are
depicted versus position $\xi$ for different values of the positron-to-cool electron density ratio $\beta$. We have taken:
$\beta=0.0$ (solid curve), $0.005$ (dashed curve), and $0.01$ (dot-dashed
curve). The other parameter values are: $\alpha=1$, $\sigma=\theta=0.01$,
$\kappa=4.0$ and $M=1.1$.}
\label{fig5}
\end{center}
\end{figure}

\begin{figure}
\begin{center}
\includegraphics[height=5.4in]{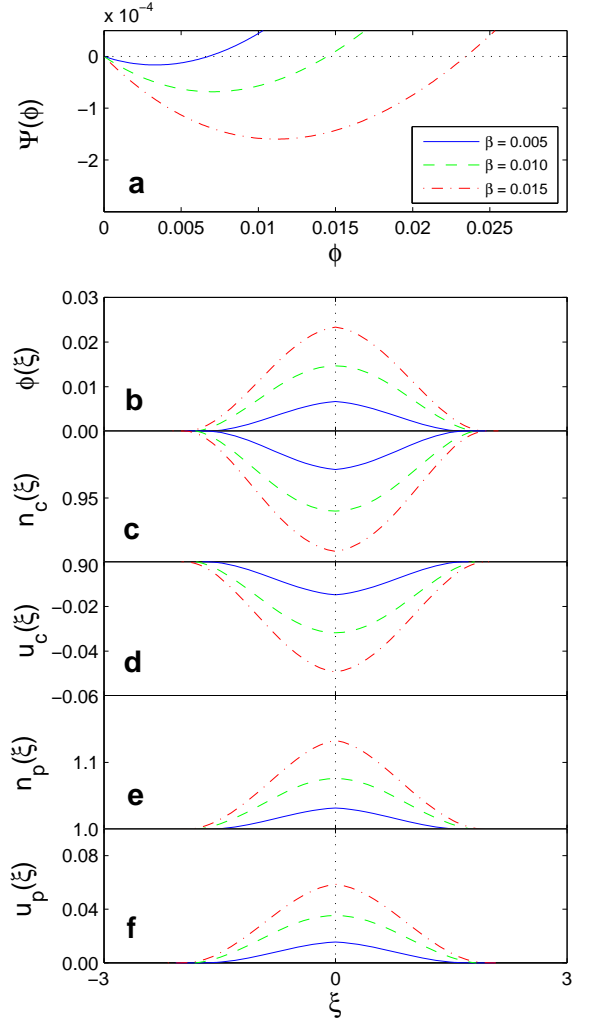}
\caption{(a) The pseudopotential $\Psi(\phi)$ of positive
polarity electrostatic solitons and the associated solutions: (b) electric potential
pulse $\phi$, (c) density $n_{c}$ and (d) velocity $u_{c}$ of the cool electron fluid,
and (e) density $n_{p}$ and (f) velocity $u_{p}$ of the positron fluid are
depicted versus position $\xi$ for different values of the positron-to-cool electron density ratio $\beta$. We have taken: $\beta=0.005$ (solid curve), $0.010$ (dashed curve),
and $0.015$ (dot-dashed curve). The other parameter values are: $\alpha=1$,
$\sigma=\theta=0.01$, $\kappa=4.0$ and $M=0.5$.}
\label{fig6}
\end{center}
\end{figure}

Figure \ref{fig5}(a) shows the variation of the pseudopotential $\Psi(\phi)$
of negative polarity solitons with the normalized negative potential $\phi$, for
different values of the positron-to-cool electron density ratio $\beta$
(keeping $\alpha=1$, $\sigma=\theta=0.01$, $\kappa=4.0$ and Mach number
$M=1.1$, all fixed). The electrostatic pulse $\phi$ shown in Fig.
\ref{fig5}(b) is obtained via a numerical integration. The negative pulse
amplitude decreases with increasing $\beta$. We algebraically determined the
fluid density (Fig. \ref{fig5}c) and velocity disturbance (Fig. \ref{fig5}d)
of the cool electrons, as well as the fluid density (Fig. \ref{fig5}e) and
velocity disturbance (Fig. \ref{fig5}f) of the positrons. It is found that an
increase in the positron-to-cool electron density ratio $\beta$ decreases the disturbances and amplitudes of $n_{c}$, $u_{c}$,
$n_{p}$ and $u_{p}$ in the negative polarity electrostatic mode. This means that
increasing the positron density reduces the negative potential solitary waves, 
in agreement with the previous results \cite{Jilani2014}. 
We also note that the profiles become less steeper but broader.

\begin{figure}
\begin{center}
\includegraphics[height=4.3in]{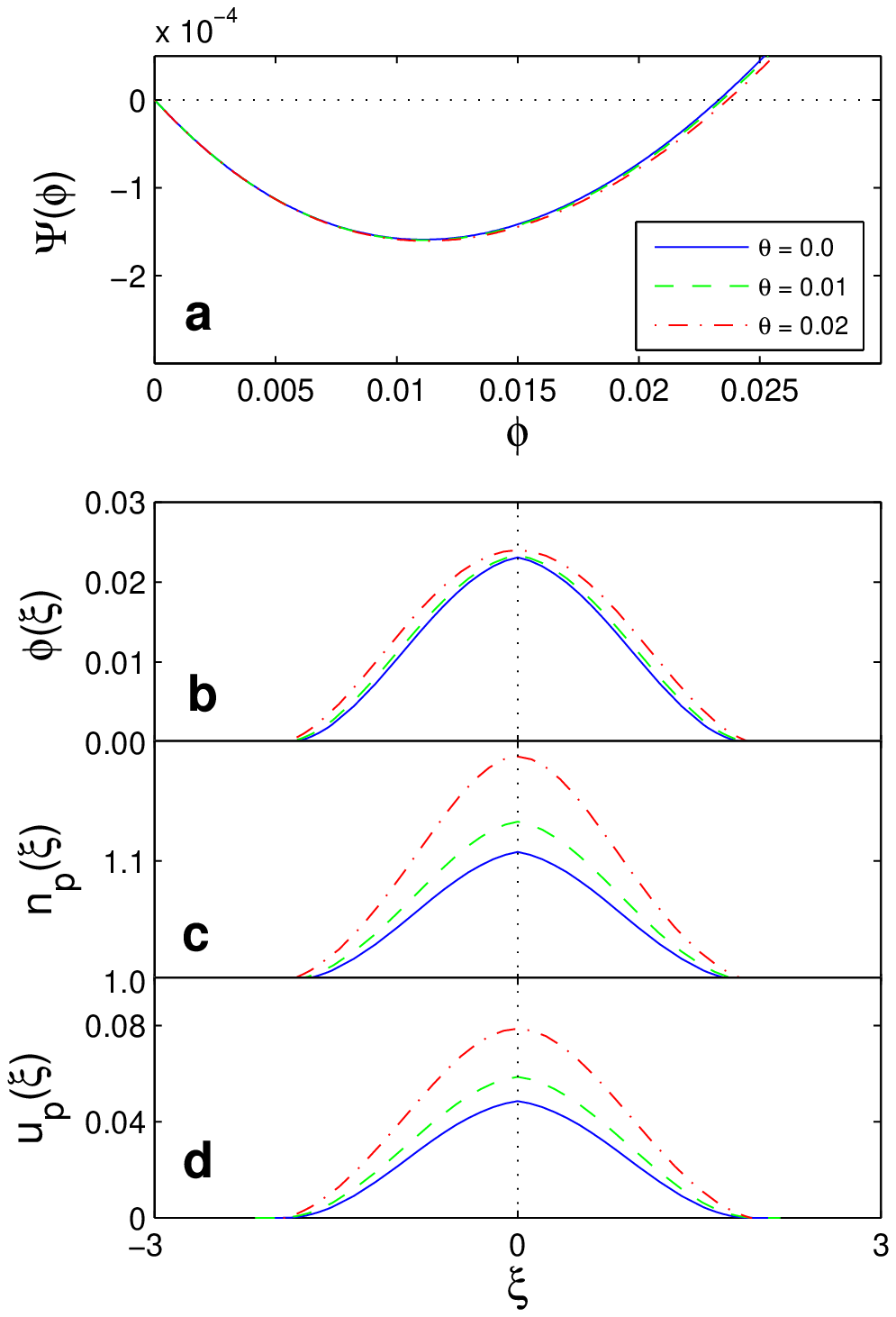}
\caption{(a) The pseudopotential $\Psi(\phi)$ of positive
polarity electrostatic solitons and the associated solutions: (b) electric potential
pulse $\phi$, and (c) density $n_{p}$ and (d) velocity $u_{p}$ of the positron
fluid are depicted versus position $\xi$ for different values of 
the positron-to-hot electron temperature ratio $\theta$. We have taken: $\theta=0.0$ (solid curve), $0.01$
(dashed curve), and $0.02$ (dot-dashed curve). The other parameter values are:
$\alpha=1$, $\sigma=0.01$, $\beta=0.015$, $\kappa=4.0$ and $M=0.5$.}
\label{fig7}
\end{center}
\end{figure}

Similarly, Figure \ref{fig6}(a) depicts the variation of the pseudopotential
$\Psi(\phi)$ of positive polarity solitons associated with the positrons for different values of the positron-to-cool
electron density ratio $\beta$ (keeping $\alpha=1$, $\sigma=\theta=0.01$, 
$\kappa=4.0$ and Mach number $M=0.5$, all fixed). As seen in Fig.~\ref{fig6}(b), that the
positive pulse amplitude rises with an increase in $\beta$, in contrast to
what we see in Fig.~\ref{fig5}(b). Furthermore, an increase in $\beta$
increases the disturbances, amplitudes and steepness of $n_{c}$ $u_{c}$, $n_{p}$ and $u_{p}$ in
the positive polarity electrostatic mode.  This means that increasing the positron
density increases the positive potential solitary waves, which agrees with the results of Ref.~\onlinecite{Jilani2014} (they used $n_{p,0}/n_{h,0}$ rather than $\beta=n_{p,0}/n_{c,0}$).

\begin{figure}
\begin{center}
\includegraphics[height=4.31in]{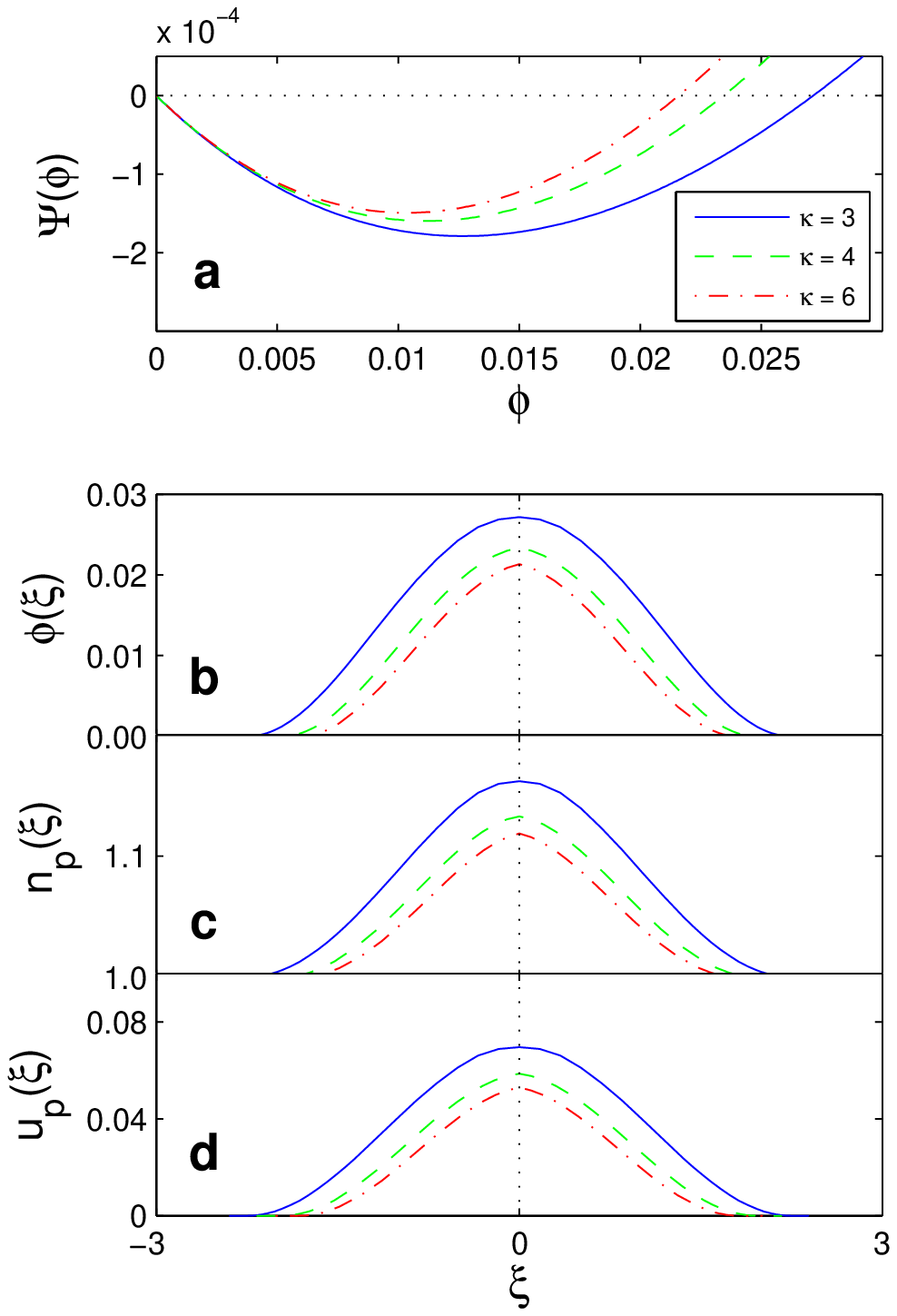}
\caption{(a) The pseudopotential $\Psi(\phi)$ of positive
polarity electrostatic solitons and the associated solutions: (b) electric potential
pulse $\phi$, and (c) density $n_{p}$ and (d) velocity $u_{p}$ of the positron
fluid are depicted versus position $\xi$, for different values of the spectral index $\kappa$. We have
taken: $\kappa=3$ (solid curve), $4$ (dashed curve), and $6$ (dot-dashed
curve). The other parameter values are: $\alpha=1$, $\sigma=\theta=0.01$,
$\beta=0.015$ and $M=0.5$.}
\label{fig8}
\end{center}
\end{figure}

The thermal effect of the positrons through $\theta=T_{p}/T_{h}$ is shown in
Fig. \ref{fig7}. The soliton excitation $\phi$ is slightly amplified  with an
increase in the temperature ratio $\theta$, which agrees with the results of Ref.~\onlinecite{Jilani2014} (they used $T_{h}/T_{p}$ rather than $\theta=T_{p}/T_{h}$). 
Furthermore, an increase in
$\theta$ slightly increases the disturbance of $n_{c}$ and $u_{c}$ (not shown here),
however, significantly increases and steepens the disturbance of $n_{p}$ and $u_{p}$ in
the positive polarity electrostatic mode (Fig. \ref{fig7}(c) and (d)). The temperature ratio $\theta$ does not make
a significant contribution to the negative polarity electrostatic solitary waves due to
the small value of $\beta$.

\begin{figure}
\begin{center}
\includegraphics[width=3.3in]{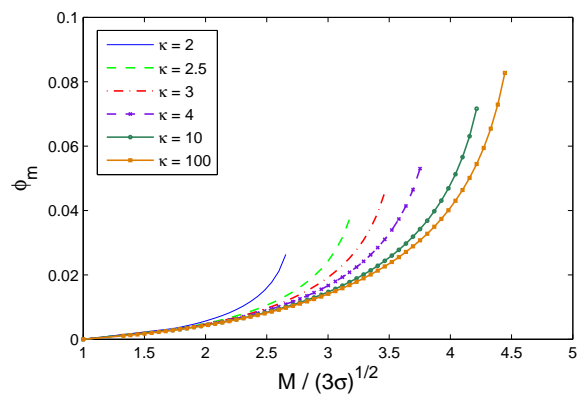}
\caption{The dependence of the pulse amplitude $\phi_{m}$ of
positive polarity electrostatic solitons on the Mach number-to-sound-speed ratio
$M/\sqrt{3{\sigma}}$ is depicted, for different values of the spectral index $\kappa$. From top
to bottom: $k=2$\ (solid curve); $2.5$\ (dashed curve); $3$\ (dot-dashed
curve); $4$\ (crosses); $10$\ (circles); $100$\ (solid squares). Here,
$\alpha=1$ and $\sigma=\theta=\beta=0.01$. }
\label{fig9}
\end{center}
\end{figure}

Figure \ref{fig8}(a) shows the pseudopotential $\Psi(\phi)$ of positive 
polarity solitons for different values of the spectral index $\kappa$ (keeping $\alpha=1$, $\sigma=\theta=0.01$, 
$\beta=0.015$ and Mach number $M=0.5$). The positive polarity electrostatic pulse
shown in Fig.~\ref{fig8}(b) is found to increase for lower $\kappa$, implying
an amplification of the electric potential disturbance as the suprathermality
increases. It can be seen that the positron fluid density (Fig.~\ref{fig8}(c))
and velocity disturbance (Fig.~\ref{fig8}(d)) are increased in the positive
polarity electrostatic mode, and again, for lower $\kappa$ values.

As inherently super-acoustic solitons are taken, it is important to see the effect of a varying true Mach number, so we explore the pulse amplitude $\phi_{m}$ of
the positive polarity electrostatic solitons as a function of the propagation speed $M$, measured relative to the true acoustic speed, $M_{1}$. The variation of the soliton amplitude $\phi_{m}$ 
as a function of the true Mach number, $M/M_{1}$, is numerically obtained from Eq.~(\ref{eq_43}). Noting that the lower limit ($M_1$) for positive polarity solitary structures is about $\sqrt{3{\sigma}}$, we have plotted the soliton amplitude $\phi_{\mathrm{m}}$ against the ratio $M/\sqrt{3{\sigma}}$, for a range of values of the parameter $\kappa$ in Fig.~\ref{fig9}. It is seen that the soliton amplitude $\phi_{m}$ increases with $M/\sqrt{3{\sigma}}$ for all values of $\kappa$. Moreover, the soliton amplitude increases with growing the suprathermality (reducing $\kappa$) at a fixed true Mach number, $M/\sqrt{3{\sigma}}$, in contrast to the results obtained previously \cite{Danehkar2011}. However, the maximum value of soliton amplitude is found to be for a Maxwellian distribution ($\kappa\rightarrow\infty$) at larger true Mach numbers ($M/\sqrt{3{\sigma}}>4$).

\section{Conclusions\label{sec:conclusion}}

In this work, we have investigated nonlinear characteristics of
electrostatic solitary wave structures in a collisionless plasma
consisting of adiabatic cool electrons, mobile cool positrons (electron holes), hot $\kappa
$-distributed electrons and immobile ions. We have derived a linear dispersion relation, and studied the effects of
positron parameters on the dispersion characteristics, through the
positron-to-cool electron density ratio $\beta$. It is found that the phase
speed increases weakly with an increase in $\beta$ (see Fig.~\ref{fig1}). Similarly, in agreement with the previous finding ~\cite{Danehkar2011}, increasing suprathermality (decreasing $\kappa$) significantly reduces the phase speed.

The Sagdeev's pseudopotential technique was used to determine nonlinear structures
and the range of allowed Mach numbers of electrostatic solitons. The results of this study indicate that
increasing the positron-to-cool electron density ratio $\beta$ and the
positron-to-hot electron temperature ratio $\theta$ lead to a slightly
narrowing of the Mach number range for negative polarity solitons (Fig.~\ref{fig2}).
Moreover, the upper Mach number limit for positive polarity solitons slightly decreases
with increasing $\beta$ and decreasing $\theta$ (Fig.~\ref{fig4}). However, the lower Mach number limit for positive polarity solitons is found to be at about $\sqrt{3{\sigma}}$ in the parameter ranges of the positrons ($\beta\lesssim0.06$ and $\theta\ll0.1$). From Fig.~\ref{fig4} one can see that increasing $\kappa$ toward a Maxwellian distribution increases the upper limit of Mach numbers for positive polarity solitons.

The e-p model predicts the existence of positive potential
solitons associated with the positrons, in addition to negative potential solitons. It is found that
increasing the positron-to-cool electron density ratio $\beta$ decreases the
normalized negative potential (Fig.~\ref{fig5}), and increases the normalized positive potential
(Fig.~\ref{fig6}) in the ranges of allowed Mach numbers for negative and positive
polarity solitons, respectively. The disturbances and amplitudes of cool electron
density and cool electron velocity due to the solitary waves decrease with
increasing $\beta$, as well as the disturbances and amplitudes of positron
density and velocity decrease in the negative polarity electrostatic mode (Fig.~\ref{fig5}). However,
higher $\beta$ increases and steepens the normalized positive potential, the
disturbances and amplitudes of positron density and velocity, and
cool electron density and velocity in the
positive polarity electrostatic mode (Fig.~\ref{fig6}). Therefore, increasing the positron density
increases the electric potential amplitude in the positive polarity electrostatic mode,
whereas decreases it in the negative polarity electrostatic (electron-acoustic) mode.

We also note that at fixed values of the normalized soliton
speed, $M$, the amplitudes of the perturbations of positron density and
velocity are significantly increased and steepened with higher values of the positron-to-hot electron
temperature ratio $\theta$ (Fig.~\ref{fig7}). As the positrons constitute a small
fraction of the total number density ($\beta\lesssim0.06$), the normalized potential,
cool electron density and velocity are trivially affected by $\theta$.
Therefore, thermal effects of the cool positrons are negligible for both
negative and positive electric potentials.

From Fig.~\ref{fig8}, it can be seen that the suprathermality can
significantly raise the electric potential amplitude in the positive
polarity electrostatic mode. This means that suprathermal electrons play a key role in
rising a positive potential pulse from a tiny fraction of cool positrons.
Therefore, we expect to have a strong positive polarity electrostatic wave when the
suprathermality is stronger (lower $\kappa$). Fig.~\ref{fig9} demonstrates how 
the pulse amplitude of positive polarity electrostatic solitons rises with reducing the spectral index
$\kappa$ (higher suprathermality) at a fixed true Mach number ($M/\sqrt{3{\sigma}}$), 
while the soliton amplitude increases with the true Mach number for 
all values of $\kappa$.

In conclusion, the results of this study suggest that the dynamics of
electrostatic solitary waves can be modified by a small fraction of cool
positrons (or electron holes) in the presence of suprathermal electrons. The results of this study
could have important implications for positive polarity electrostatic waves
 observed in the auroral magnetosphere
\cite{Muschietti1999,Cattell2002}, as well as the formation of coherent radio emission in pulsars \cite{Sturrock1971,Ruderman1975,Arons1979,Arons1983,Melikidze2000}, 
where positrons and suprathermal
electrons are present.

\section*{Acknowledgements}

The author acknowledges the award of a postgraduate studentship from the Department for Employment and
Learning (DEL) in Northern Ireland at the Queen's University Belfast, and a Research Excellence Scholarship from Macquarie University. 

%\bibliographystyle{aipnum4-1}
%\bibliography{references}
% Produces the bibliography via BibTeX.

%

\end{document}